\documentclass[aip,reprint,twocolumn,groupedaddress, eprintnumbers]{revtex4-2}

\usepackage{chemformula} 
\usepackage[]{graphicx}
\usepackage{epstopdf}
\usepackage{bm}
\usepackage{MnSymbol}
\usepackage{latexsym}
\usepackage{amsmath}
\usepackage{physics}
\usepackage{natbib}

\begin{document}

\title{Dissipation-Induced Symmetry Breaking: Emphanitic Transitions in Lead- and Tin-Containing Chalcogenides and Halide Perovskites}

\author{Kingshuk Mukhuti}\email{kingshukmukhuti.phy@gmail.com}
\affiliation{Indian Institute of Science Education \& Research Kolkata, Mohanpur, Nadia 741246, West Bengal, India}
\author{Sudip Sinha}\email{ss19rs063@iiserkol.ac.in}
\affiliation{Indian Institute of Science Education \& Research Kolkata, Mohanpur, Nadia 741246, West Bengal, India}
\author{Subhasis Sinha}\email{subhasis@iiserkol.ac.in}
\affiliation{Indian Institute of Science Education \& Research Kolkata, Mohanpur, Nadia 741246, West Bengal, India}
\author{Bhavtosh Bansal}\email{bhavtosh@iiserkol.ac.in}
\affiliation{Indian Institute of Science Education \& Research Kolkata, Mohanpur, Nadia 741246, West Bengal, India}

\begin{abstract}
Lead and tin-based chalcogenide semiconductors like PbTe or SnSe have long been known to exhibit an unusually low thermal conductivity that makes them very attractive thermoelectric materials. An apparently unrelated fact is that the excitonic bandgap in these materials increases with temperature, whereas for most semiconductors one observes the opposite trend. These two anomalous features are also seen in a very different class of photovoltaic materials, namely the halide-perovskites such as CsPbBr$_3$. It has been previously proposed that {\em emphanisis}, a local symmetry-breaking phenomenon, is the one common origin of these unusual features. Discovered a decade ago, emphanisis is the name given to the observed displacement of the lead or the tin ions from their cubic symmetry ground state to a locally distorted phase at high temperature. This phenomenon has been puzzling because it is unusual for the high-temperature state to be of a lower symmetry than the degenerate ground state. Motivated by the celebrated vibration-inversion resonance of the ammonia molecule, we propose a quantum tunneling-based model for emphanisis where decoherence is responsible for the local symmetry breaking with increasing temperature. From the analytic expression of the temperature dependence of the tunnel splitting (which serves as an order parameter), we provide three-parameter fitting formulae which capture the observed temperature dependence of the ionic displacements as well as the anomalous increase of the excitonic bandgap in all the relevant materials.
\end{abstract}
\maketitle
The BY-type chalcogenides (B=Sn, Pb and Y=Te, S, Se) with the rock-salt structure\cite{Bozin_Science} and the ABX-type  (A=Cs, methyl-ammonium, etc., B=Pb, Sn, and X=Cl, Br, I) halide perovskites\cite{Fabini_JACS} are two classes of Pb or Sn-containing semiconductors that have been of great interest---the former in the context of thermoelectricity\cite{Ravich} and the latter as photovoltaics.\cite{Huang_PerovskiteReview, Brenner} Interestingly, these two rather disparate material systems share some unusual features---a very low thermal conductivity\cite{Zeier, Xie-Kanatzidis} that enhances the thermoelectric figure-of-merit and an anomalous increase of the excitonic bandgap with temperature.\cite{Yang,Yu,Saran,Handa,Gibson,Gibbs,Paul,Tauber} It has been proposed that these anomalies arise on account of a symmetry-breaking phenomenon, named emphanisis.\cite{Bozin_Science, Fabini_JACS, Fabini_MRS, Kontos} The lead or the tin ion, which is common to all these materials, sits at a lattice site with cubic symmetry in the low temperature ground state. At around $100$ K, a puzzling sub-angstrom displacement has been inferred from the atomic pair distribution function (PDF) measurements.\cite{Bozin_Science, Fabini_JACS} The distortion is local and does not lead to any macroscopic order.\cite{Aeppli, Fabini_JACS, Fabini_MRS,Bozin_Science}

In this Letter, we propose that emphanisis be understood in analogy with the vibration-inversion ``umbrella'' resonance of the ammonia molecule.\cite{Lehn} The lone-pair-containing lead or tin ions, through the process of quantum tunneling between the equivalent elastic potential energy minima, form coherently superposed extended states with a zero mean displacement. The environment-induced dephasing\cite{Silbey_Harris,Leggett_RMP,Zurek,Joos1996} at higher temperatures destroys this superposition, causing the emphanitic crossover.

\begin{figure}[!b]
    \includegraphics[scale=0.17]{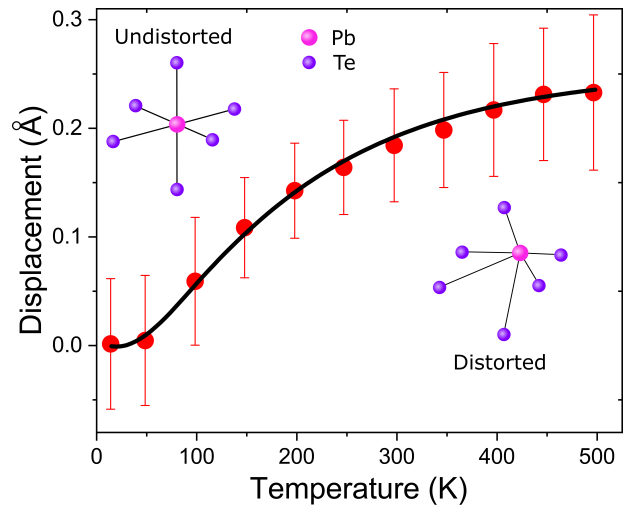}
	\caption{Emphanisis: Data points (with error bars) are taken from the first observation by Bo{\v z}in, et al. \cite{Bozin_Science} of the temperature-induced off-centering of the Pb ions in PbTe, inferred from the synchrotron-based PDF measurement. Solid line is the fit to equation (3), with $\lambda=1$, $d_\infty=0.253$ \AA, $T_0 = 158$ K, where $T_0=\hbar\Omega_0/k_B$.}
	\label{Figure 1}
\end{figure}
\begin{figure}[!b]
	\includegraphics[scale=0.19]{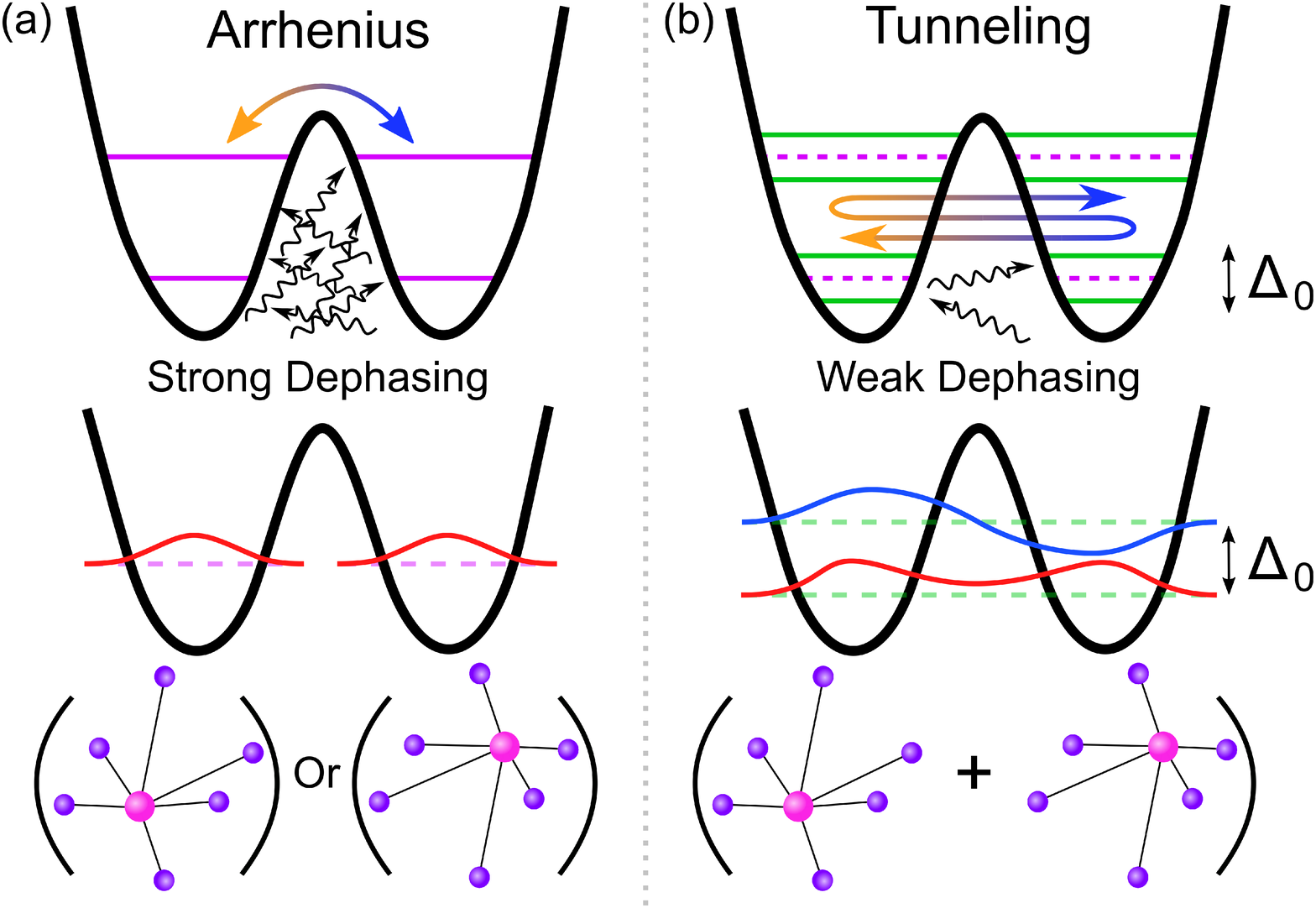}
	\caption{Depiction of the (a) classical,\cite{Fabini_JACS} and (b) the quantum mechanical tunneling description. In both cases, the lead or the tin ion is in a landscape of two (or more) equivalent potential energy minima corresponding to off-centered low-energy positions. In the classical picture, the ion's location is well defined due to the environment-induced scattering events. The transition to the other minima can only occur via Arrhenius barrier-crossing, where the barrier height is much larger than bath temperature. In the tunneling picture, the energetically degenerate wave functions of the minima hybridize to form extended bonding and antibonding states. These states are separated by a tunnel gap, $\Delta_r(T)$. The emphanisis corresponds to the dissipation-induced renormalization of this gap to zero.}
	\label{Figure 2}
\end{figure}

Significantly, like the nitrogen ion\cite{Lehn} in NH$_3$, the $B$ (= Sn or Pb) ion [Figure \ref{Figure 1}] in the above materials  has a lone pair of electrons responsible for the stabilization of the distorted state.\cite{Fabini_JACS, Fabini_MRS} The displacement of the $B$ ion amounts to its moving to one of the two (or many) energetically degenerate lower symmetry locations, depicted for simplicity as a symmetric double-well potential\cite{Fabini_JACS} in Figure \ref{Figure 2}. Classically [Figure \ref{Figure 2} (a)]\cite{Fabini_JACS}, one expects the low temperature ground state to have broken symmetry as the ion is trapped into one of the minima on account of insufficient thermal energy. At high temperatures, this symmetry could possibly be restored in the classical picture through an emergent time-averaged (ergodic) state, with the B-type ion undergoing rapid Arrhenius barrier crossings.

Let us consider an alternative picture\cite{Simonius} with two quantum mechanical ingredients---tunnelling\cite{Lehn} and dissipation.\cite{Leggett_RMP,Silbey_Harris, Wagner} On account of tunneling, the energetically degenerate classical ground states, corresponding to the distorted structures, would hybridize.\cite{Joos1996, Leggett_RMP} The entanglement due to the coherent superposition of the wave functions of the left and the right well, $|L\rangle$ and $|R\rangle$ respectively, would result in bonding  $|\Phi_+\rangle ={1\over 2}[|L\rangle+|R\rangle]$ and anti-bonding $|\Phi_-\rangle ={1\over 2} [|L\rangle-|R\rangle ]$ wave functions, with (bare) tunnel-splitting energy\cite{Leggett_RMP} $\Delta_0$. Note that the expectation value of the position $\langle x\rangle=\langle\Phi_\pm |x|\Phi_\pm\rangle$ is zero in each case, meaning that the atom would, in this average sense, be found at the high symmetry location. $\Delta_0$ is related to the tunneling time $\tau_t\approx\hbar/\Delta_0$ for the repeated back and forth {\em phase-coherent} excursions between the two minima. The absence of phase-breaking events is crucial for the success of such coherent superpositions. If the tunneling would become dissipative, viz. there are phase-breaking events with a relaxation time $\tau_{_{PB}}$, the tunnel-splitting energy $\Delta_0$ would renormalize to $\Delta_r(T)$ (vide infra), and eventually collapse to zero when $\tau_{_{PB}}<<\tau_t$. The particle would then localize in one of the wells, marking the quantum-to-classical crossover.\cite{Silbey_Harris, Wagner, Leggett_RMP, Simonius} This is emphanisis in a nutshell.

The Hamiltonian\cite{Wagner}
\begin{equation}\label{eqn: Hamiltonian}
H=\Delta_0\hat{\sigma}_x+\hbar\Omega_0 \left({1\over 2}+\hat{a}^\dagger \hat{a}\right)+ \kappa\hat{\sigma}_z(\hat{a}+\hat{a}^\dagger),
\end{equation}
where a single spin is coupled to an Einstein boson, provides the minimal rigorous description for the physics described above. $\hat{\sigma}_x$ and $\hat{\sigma}_z$ are the Pauli spin matrices and $\hat{a}$ ($\hat{a}^\dagger$) is the annihilation (creation) operator for a boson with frequency $\Omega_0$ and the number expectation value $\langle \hat{a}^\dagger \hat{a}\rangle= n_B(\Omega_0,T)=[\exp(\hbar\Omega_0/k_BT)-1]^{-1}$.

This Hamiltonian describes the competition between the non-commuting operators $\hat{\sigma}_z$ and $\hat{\sigma}_x$. $\hat{\sigma}_x$  is diagonal for the delocalized states $\lbrace |\Phi_+\rangle, |\Phi_-\rangle\rbrace$  spanning the two wells. The $\hat{\sigma}_z$ operator is diagonal in the basis of the localized states $\lbrace |L\rangle,  |R\rangle\rbrace$. If the system starts as a delocalized state, the environment manifesting as $\hat{\sigma}_z$ would attempt ``measurements'' causing localization into one of the two wells. After this collapse, the system would again attempt to evolve and tunnel out (on account of $\hat{\sigma}_x$ in $H$) but if the time between such successive measurements is smaller than the characteristic tunneling time, these attempts become Sisyphean tasks. The system would cease to evolve (quantum Zeno effect)\cite{Joos1996} and localize. Coupling of $\hat{\sigma}_z$ to the bosonic excitation $\hat{a}$ implies that the effect of the environment becomes stronger with the increase in temperature. Dissipation thus superselects\cite{Zurek} one of the two minima, causing the emphanitic crossover. An infinite tunneling time would naturally correspond to the complete collapse of the tunneling gap. Therefore it is sufficient to think of the renromalized gap $\Delta_r(T)$ as the ``order parameter'' for this crossover. In the limit $\Delta_0 <<k_BT$,
\begin{equation}\label{Eqn:SplittingFinal}
\Delta_r(T)\approx \tilde{\Delta}_0\exp[-\lambda n_B(\Omega_0,T)].
\end{equation}
This deceptively simple closed form expression is derived (see {\it Supplementary Material} at the end of the manuscript) using the  unitary transformation-based variational method.\cite{Silbey_Harris, Wagner} $\lambda\equiv 2\left({\kappa\over \hbar\Omega_0}\right)^2$ and $\tilde{\Delta}_0\equiv \Delta_0\exp[-\lambda/2]$ is the tunnel gap at zero temperature, so defined because the zero-point renormalization is usually not discernable. Note that $\lim_{T\rightarrow \infty} \Delta_r(T)=0$, which is what we set out to have.

\begin{figure*}[t!]
	\includegraphics[scale=0.31]{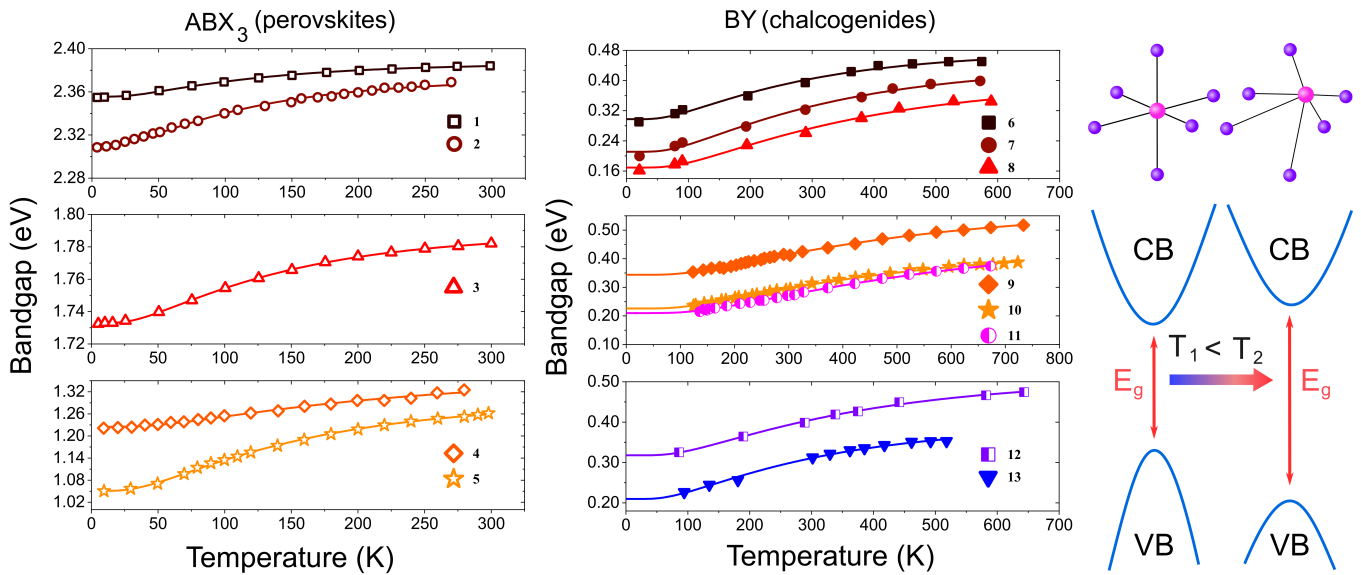}
	\centering
	\caption{Temperature dependence of the excitonic bandgap of some published reports on halide perovskites and chalcogenides along with the fits to equation (\ref{Eqn:Gap}).  The schematic on the right, adapted from Kontos, et al.,\cite{Kontos} highlights the strong correlation between emphanisis and bandgap widening.
		The legend, with the numbers in the parenthesis denoting the fitted values ($E_g^{\infty}$ in eV, $\Xi_0$ in meV, $T_{0}$ in K)  respectively---1. CsPbBr$_3$ nanocrystals (2.386, 31, 97),\cite{Saran} 2. CsPbBr$_3$ (2.371, 60, 93),\cite{Yang} 3. CsPbI$_3$ nanocrystals (1.788, 55, 111),\cite{Saran} 4. CsSnI$_3$ thin film (1.315, 95, 143),\cite{Yu} 5. MASnI$_3$ thin film (1.292, 240, 122),\cite{Handa} 6. PbS (0.477, 179, 221),\cite{Gibson} 7. PbTe (0.438, 227, 252),\cite{Gibson} 8. PbSe (0.397,228, 290),\cite{Gibson} 9. PbS (0.574, 230, 391),\cite{Gibbs} 10. PbTe (0.438, 212, 374),\cite{Gibbs} 11. PbSe (0.472, 262, 462),\cite{Gibbs} 12. PbS (0.507, 190, 287),\cite{Paul} 13. PbTe (0.391, 181, 240).\cite{Tauber} }
	\label{Figure 3}
\end{figure*}
The emphanitic crossover is thus characterized by the parameters $\tilde{\Delta}_0$, $\lambda$, and $\Omega_0$. $\lambda$, the effective system-bath coupling, manifests as the steepness and  $\hbar\Omega_0$ sets the temperature scale for the crossover. The bare tunnel splitting $\Delta_0$ may also have a strong temperature dependence from other extraneous factors, like an overall thermal expansion of the system. These may further aid emphanisis, but are ignored here.

To connect this high temperature collapse of the tunnel splitting to emphanisis [Figure \ref{Figure 1}], we assume a linear relationship between the magnitude of the local distortion $d(T)$ at temperature $T$ and $\Delta_r$, viz.,
\begin{equation}\label{Eqn: emphanisis displacement}
{d_\infty-d(T)\over d_\infty}={\Delta_r (T)\over \tilde{\Delta}_0}=\exp[-\lambda n_B(\Omega_0,T)].
\end{equation}
Here $d_\infty$ is the asymptotic $T\rightarrow \infty$ value of $d(T)$.  A satisfactory fit to Bo{\v z}in, et al.'s experimental data\cite{Bozin_Science} is obtained [Figure \ref{Figure 1}] with equation (\ref{Eqn: emphanisis displacement}), even if we fix $\lambda=1$ to keep the number of fitting parameters to an absolute minimum.

If $T_0$ ($= \hbar\Omega_0/k_B$) is the characteristic temperature for emphanisis, then the condition $\Delta_0<<k_BT_0$ must necessarily hold because $\Delta_0$ is in the microwave frequency range corresponding to sub-kelvin temperature.\cite{Prager-Heidemann} We emphasize that this difference in the energy scales does not fundamentally constrain the coherence because it is not the bath temperature, but the effectiveness of the coupling to environment (indicated by the linewidth of the resonance) that is important. One need only recall that radio-frequency nuclear magnetic resonance is routinely resolved at room temperature.

Since the early 1970s, atom tunneling has been identified as an essential ingredient for understanding a diverse range of phenomena, including the low temperature specific heat anomalies in amorphous materials and many chemical reactions up to tens of kelvin.\cite{Ley_Gerbig_Schreiner, Benderskii} Tunnel splitting of the rotational, librational and vibration-inversion states is also responsible for microwave resonances in a large number of molecules,\cite{Prager-Heidemann} even well above room temperature.
In the gas phase, pressure (collision) broadening is the primary mechanism connected with the disappearance of such resonance lines.\cite{Joos1996} In particular, the potentially disastrous radiation broadening has been found to be of negligible importance, even up to high temperatures.\cite{Joos1996} The lead or the tin ions are relatively well protected in the locally crystalline environment because electronic scattering (due to the large mass difference) would not be effective in destroying the coherence. While understanding the precise mechanism for decoherence is both subtle\cite{Leggett_RMP} and would be presently untenable on account of the unknown scattering cross sections\cite{Joos1996} for our problem, we are effectively arguing that the modes most efficient in dephasing are gapped and only build up in a sufficient density around the characteristic temperature $T_0(= \hbar\Omega_0/k_B)$. For the proposed mechanism to be sensible, it is also crucial that emphanisis be a local phenomenon without any macroscopic order as the tunnel splitting would be exponentially suppressed with the coherence size.

Finally, and most remarkably, such atom-tunneling (with the tunnel splitting  of $0.46$ meV) has actually been reported right up to room temperature in a very recent inelastic neutron scattering study on BaTiS$_3$.\cite{sun2020NatComm} Although the study draws very different conclusions from ours, it nevertheless does seem to vindicate our premise that such resonances may persist above 100 K.

\noindent
{\em Anomalous temperature dependence of the bandgap.---}In conventional band insulators and semiconductors, the bandgap reduces with the increase in temperature and follows the total phonon density.\cite{Lautenschlager_Cardona} Contrarily, the lead- and tin-based halide perovskites and the chalcogenides exhibit (i) an anomalous {\em increase} of the excitonic bandgap $E_g(T)$ with temperature, and (ii) a non-monotonic rate of change of the gap with temperature---$\partial E_g(T)/\partial T$ has a distinct peak at an intermediate temperature ($\sim 100 $ K) with a clear correlation of the gap with emphanisis\cite{Fabini_JACS, Kontos} (compare Figure \ref{Figure 1} and Figure \ref{Figure 3}).
Based on arguments analogous to those made for introducing equation (\ref{Eqn: emphanisis displacement}) above, we may expect
\begin{equation}\label{Eqn:Gap}
E_g(T)=E_g^\infty-\Xi_0\exp[-\lbrace\exp(T_{0}/T)-1\rbrace ^{-1}].
\end{equation}
$E_g^\infty$ is the post-emphanisis high-temperature bandgap, $T_{0}$ the characteristic phonon temperature, $(E_g^\infty-\Xi_0)$ is the gap at zero temperature, and therefore $|\Xi_0|$ is magnitude of the bandgap change affected by the emphanisis. To keep the fitting parameters to the minimum, we have further set the constant equivalent to $\lambda$ in equation (\ref{Eqn:SplittingFinal}) to unity. The temperature dependence of the absorption edge of various halide perovskites\cite{Yang, Yu,Saran, Handa} and chalcogenides,\cite{Gibson, Gibbs, Paul, Tauber} fitted to  equation (\ref{Eqn:Gap}), is shown in Figure \ref{Figure 3}.

We must mention that in using the simplest of the spin-boson models and a na\"ive version of the lone-pair stereochemistry,  we have sidestepped important details regarding the nature of the potential wells, anomalies in the phonon dispersion, lattice softening and expansion, the precise relationship between the dynamic distortion and the bandgap, features of the electronic band structure like inversion and degeneracy, etc. These specific factors must indeed be incorporated for any meaningful materials engineering. Nevertheless, the excellent fits in Figure \ref{Figure 1} and Figure \ref{Figure 3} do seem to testify to the validity of the general thesis we have proposed here---tunneling and environment superselection are the primary cause for emphanisis.



See the Supplementary Material for the derivation of Eq. (2) and fits by Eq. (4) to more data on the temperature dependence of the bandgap in lead-containing halide perovskites.

We thank Siddhartha Lal for many useful comments. BB thanks Science
and Engineering Research Board, Department of Science and Technology, Government of India,
for the Core Research Grant (CRG/2018/003282).
%

\bibliography{arxiv_DissipativeTunneling}

\begin{thebibliography}{35}%
\makeatletter
\providecommand \@ifxundefined [1]{%
 \@ifx{#1\undefined}
}%
\providecommand \@ifnum [1]{%
 \ifnum #1\expandafter \@firstoftwo
 \else \expandafter \@secondoftwo
 \fi
}%
\providecommand \@ifx [1]{%
 \ifx #1\expandafter \@firstoftwo
 \else \expandafter \@secondoftwo
 \fi
}%
\providecommand \natexlab [1]{#1}%
\providecommand \enquote  [1]{``#1''}%
\providecommand \bibnamefont  [1]{#1}%
\providecommand \bibfnamefont [1]{#1}%
\providecommand \citenamefont [1]{#1}%
\providecommand \href@noop [0]{\@secondoftwo}%
\providecommand \href [0]{\begingroup \@sanitize@url \@href}%
\providecommand \@href[1]{\@@startlink{#1}\@@href}%
\providecommand \@@href[1]{\endgroup#1\@@endlink}%
\providecommand \@sanitize@url [0]{\catcode `\\12\catcode `\$12\catcode
  `\&12\catcode `\#12\catcode `\^12\catcode `\_12\catcode `\%12\relax}%
\providecommand \@@startlink[1]{}%
\providecommand \@@endlink[0]{}%
\providecommand \url  [0]{\begingroup\@sanitize@url \@url }%
\providecommand \@url [1]{\endgroup\@href {#1}{\urlprefix }}%
\providecommand \urlprefix  [0]{URL }%
\providecommand \Eprint [0]{\href }%
\providecommand \doibase [0]{https://doi.org/}%
\providecommand \selectlanguage [0]{\@gobble}%
\providecommand \bibinfo  [0]{\@secondoftwo}%
\providecommand \bibfield  [0]{\@secondoftwo}%
\providecommand \translation [1]{[#1]}%
\providecommand \BibitemOpen [0]{}%
\providecommand \bibitemStop [0]{}%
\providecommand \bibitemNoStop [0]{.\EOS\space}%
\providecommand \EOS [0]{\spacefactor3000\relax}%
\providecommand \BibitemShut  [1]{\csname bibitem#1\endcsname}%
\let\auto@bib@innerbib\@empty
\bibitem [{\citenamefont {Bo{\v z}in}\ \emph {et~al.}(2010)\citenamefont {Bo{\v
  z}in}, \citenamefont {Malliakas}, \citenamefont {Souvatzis}, \citenamefont
  {Proffen}, \citenamefont {Spaldin}, \citenamefont {Kanatzidis},\ and\
  \citenamefont {Billinge}}]{Bozin_Science}%
  \BibitemOpen
  \bibfield  {author} {\bibinfo {author} {\bibfnamefont {E.~S.}\ \bibnamefont
  {Bo{\v z}in}}, \bibinfo {author} {\bibfnamefont {C.~D.}\ \bibnamefont
  {Malliakas}}, \bibinfo {author} {\bibfnamefont {P.}~\bibnamefont
  {Souvatzis}}, \bibinfo {author} {\bibfnamefont {T.}~\bibnamefont {Proffen}},
  \bibinfo {author} {\bibfnamefont {N.~A.}\ \bibnamefont {Spaldin}}, \bibinfo
  {author} {\bibfnamefont {M.~G.}\ \bibnamefont {Kanatzidis}},\ and\ \bibinfo
  {author} {\bibfnamefont {S.~J.~L.}\ \bibnamefont {Billinge}},\ }\bibfield
  {title} {\enquote {\bibinfo {title} {Entropically stabilized local dipole
  formation in lead chalcogenides},}\ }\href
  {https://doi.org/10.1126/science.1192759} {\bibfield  {journal} {\bibinfo
  {journal} {Science}\ }\textbf {\bibinfo {volume} {330}},\ \bibinfo {pages}
  {1660--1663} (\bibinfo {year} {2010})}\BibitemShut {NoStop}%
\bibitem [{\citenamefont {Fabini}\ \emph {et~al.}(2016)\citenamefont {Fabini},
  \citenamefont {Laurita}, \citenamefont {Bechtel}, \citenamefont {Stoumpos},
  \citenamefont {Evans}, \citenamefont {Kontos}, \citenamefont {Raptis},
  \citenamefont {Falaras}, \citenamefont {der Ven}, \citenamefont
  {Kanatzidis},\ and\ \citenamefont {Seshadri}}]{Fabini_JACS}%
  \BibitemOpen
  \bibfield  {author} {\bibinfo {author} {\bibfnamefont {D.~H.}\ \bibnamefont
  {Fabini}}, \bibinfo {author} {\bibfnamefont {G.}~\bibnamefont {Laurita}},
  \bibinfo {author} {\bibfnamefont {J.~S.}\ \bibnamefont {Bechtel}}, \bibinfo
  {author} {\bibfnamefont {C.~C.}\ \bibnamefont {Stoumpos}}, \bibinfo {author}
  {\bibfnamefont {H.~A.}\ \bibnamefont {Evans}}, \bibinfo {author}
  {\bibfnamefont {A.~G.}\ \bibnamefont {Kontos}}, \bibinfo {author}
  {\bibfnamefont {Y.~S.}\ \bibnamefont {Raptis}}, \bibinfo {author}
  {\bibfnamefont {P.}~\bibnamefont {Falaras}}, \bibinfo {author} {\bibfnamefont
  {A.~V.}\ \bibnamefont {der Ven}}, \bibinfo {author} {\bibfnamefont {M.~G.}\
  \bibnamefont {Kanatzidis}},\ and\ \bibinfo {author} {\bibfnamefont
  {R.}~\bibnamefont {Seshadri}},\ }\bibfield  {title} {\enquote {\bibinfo
  {title} {{Dynamic Stereochemical Activity of the Sn$^{2+}$ Lone Pair in
  Perovskite CsSnBr$_3$}},}\ }\href {https://doi.org/10.1021/jacs.6b06287}
  {\bibfield  {journal} {\bibinfo  {journal} {J. Am. Chem. Soc.}\ }\textbf
  {\bibinfo {volume} {138}},\ \bibinfo {pages} {11820--11832} (\bibinfo {year}
  {2016})}\BibitemShut {NoStop}%
\bibitem [{\citenamefont {Ravich}, \citenamefont {Efimova},\ and\ \citenamefont
  {Smirnov}(1970)}]{Ravich}%
  \BibitemOpen
  \bibfield  {author} {\bibinfo {author} {\bibfnamefont {Y.~I.}\ \bibnamefont
  {Ravich}}, \bibinfo {author} {\bibfnamefont {B.~A.}\ \bibnamefont
  {Efimova}},\ and\ \bibinfo {author} {\bibfnamefont {I.~A.}\ \bibnamefont
  {Smirnov}},\ }\href {https://doi.org/10.1007/978-1-4684-8607-0} {\emph
  {\bibinfo {title} {Semiconducting Lead Chalcogenides}}},\ edited by\ \bibinfo
  {editor} {\bibfnamefont {L.~S.}\ \bibnamefont {Stil'bans}}\ (\bibinfo
  {publisher} {Springer {US}},\ \bibinfo {year} {1970})\BibitemShut {NoStop}%
\bibitem [{\citenamefont {Huang}\ \emph {et~al.}()\citenamefont {Huang},
  \citenamefont {Kavanagh}, \citenamefont {Scanlon}, \citenamefont {Walsh},\
  and\ \citenamefont {Hoye}}]{Huang_PerovskiteReview}%
  \BibitemOpen
  \bibfield  {author} {\bibinfo {author} {\bibfnamefont {Y.-T.}\ \bibnamefont
  {Huang}}, \bibinfo {author} {\bibfnamefont {S.~R.}\ \bibnamefont {Kavanagh}},
  \bibinfo {author} {\bibfnamefont {D.~O.}\ \bibnamefont {Scanlon}}, \bibinfo
  {author} {\bibfnamefont {A.}~\bibnamefont {Walsh}},\ and\ \bibinfo {author}
  {\bibfnamefont {R.~L.~Z.}\ \bibnamefont {Hoye}},\ }\bibfield  {title}
  {\enquote {\bibinfo {title} {Perovskite-inspired materials for photovoltaics
  -- from design to devices},}\ }\href@noop {} {\bibfield  {journal} {\bibinfo
  {journal} {Preprint at https://arxiv.org/abs/2008.08959; 2020.}\ }}\Eprint
  {https://arxiv.org/abs/http://arxiv.org/abs/2008.08959v1}
  {http://arxiv.org/abs/2008.08959v1} \BibitemShut {NoStop}%
\bibitem [{\citenamefont {Brenner}\ \emph {et~al.}(2016)\citenamefont
  {Brenner}, \citenamefont {Egger}, \citenamefont {Kronik}, \citenamefont
  {Hodes},\ and\ \citenamefont {Cahen}}]{Brenner}%
  \BibitemOpen
  \bibfield  {author} {\bibinfo {author} {\bibfnamefont {T.~M.}\ \bibnamefont
  {Brenner}}, \bibinfo {author} {\bibfnamefont {D.~A.}\ \bibnamefont {Egger}},
  \bibinfo {author} {\bibfnamefont {L.}~\bibnamefont {Kronik}}, \bibinfo
  {author} {\bibfnamefont {G.}~\bibnamefont {Hodes}},\ and\ \bibinfo {author}
  {\bibfnamefont {D.}~\bibnamefont {Cahen}},\ }\bibfield  {title} {\enquote
  {\bibinfo {title} {Hybrid organic--inorganic perovskites: Low-cost
  semiconductors with intriguing charge-transport properties},}\ }\href
  {https://doi.org/10.1038/natrevmats.2015.7} {\bibfield  {journal} {\bibinfo
  {journal} {Nature Reviews Materials}\ }\textbf {\bibinfo {volume} {1}}
  (\bibinfo {year} {2016}),\ 10.1038/natrevmats.2015.7}\BibitemShut {NoStop}%
\bibitem [{\citenamefont {Zeier}\ \emph {et~al.}(2016)\citenamefont {Zeier},
  \citenamefont {Zevalkink}, \citenamefont {Gibbs}, \citenamefont {Hautier},
  \citenamefont {Kanatzidis},\ and\ \citenamefont {Snyder}}]{Zeier}%
  \BibitemOpen
  \bibfield  {author} {\bibinfo {author} {\bibfnamefont {W.~G.}\ \bibnamefont
  {Zeier}}, \bibinfo {author} {\bibfnamefont {A.}~\bibnamefont {Zevalkink}},
  \bibinfo {author} {\bibfnamefont {Z.~M.}\ \bibnamefont {Gibbs}}, \bibinfo
  {author} {\bibfnamefont {G.}~\bibnamefont {Hautier}}, \bibinfo {author}
  {\bibfnamefont {M.~G.}\ \bibnamefont {Kanatzidis}},\ and\ \bibinfo {author}
  {\bibfnamefont {G.~J.}\ \bibnamefont {Snyder}},\ }\bibfield  {title}
  {\enquote {\bibinfo {title} {Thinking like a chemist: Intuition in
  thermoelectric materials},}\ }\href {https://doi.org/10.1002/anie.201508381}
  {\bibfield  {journal} {\bibinfo  {journal} {Angew. Chem. Int. Ed.}\ }\textbf
  {\bibinfo {volume} {55}},\ \bibinfo {pages} {6826--6841} (\bibinfo {year}
  {2016})}\BibitemShut {NoStop}%
\bibitem [{\citenamefont {Xie}\ \emph {et~al.}(2020)\citenamefont {Xie},
  \citenamefont {Hao}, \citenamefont {Bao}, \citenamefont {Slade},
  \citenamefont {Snyder}, \citenamefont {Wolverton},\ and\ \citenamefont
  {Kanatzidis}}]{Xie-Kanatzidis}%
  \BibitemOpen
  \bibfield  {author} {\bibinfo {author} {\bibfnamefont {H.}~\bibnamefont
  {Xie}}, \bibinfo {author} {\bibfnamefont {S.}~\bibnamefont {Hao}}, \bibinfo
  {author} {\bibfnamefont {J.}~\bibnamefont {Bao}}, \bibinfo {author}
  {\bibfnamefont {T.~J.}\ \bibnamefont {Slade}}, \bibinfo {author}
  {\bibfnamefont {G.~J.}\ \bibnamefont {Snyder}}, \bibinfo {author}
  {\bibfnamefont {C.}~\bibnamefont {Wolverton}},\ and\ \bibinfo {author}
  {\bibfnamefont {M.~G.}\ \bibnamefont {Kanatzidis}},\ }\bibfield  {title}
  {\enquote {\bibinfo {title} {All-inorganic halide perovskites as potential
  thermoelectric materials: Dynamic cation off-centering induces ultralow
  thermal conductivity},}\ }\href {https://doi.org/10.1021/jacs.0c03427}
  {\bibfield  {journal} {\bibinfo  {journal} {J. Am. Chem. Soc.}\ }\textbf
  {\bibinfo {volume} {142}},\ \bibinfo {pages} {9553--9563} (\bibinfo {year}
  {2020})}\BibitemShut {NoStop}%
\bibitem [{\citenamefont {Yang}\ \emph {et~al.}(2017)\citenamefont {Yang},
  \citenamefont {Surrente}, \citenamefont {Galkowski}, \citenamefont {Miyata},
  \citenamefont {Portugall}, \citenamefont {Sutton}, \citenamefont
  {Haghighirad}, \citenamefont {Snaith}, \citenamefont {Maude}, \citenamefont
  {Plochocka},\ and\ \citenamefont {Nicholas}}]{Yang}%
  \BibitemOpen
  \bibfield  {author} {\bibinfo {author} {\bibfnamefont {Z.}~\bibnamefont
  {Yang}}, \bibinfo {author} {\bibfnamefont {A.}~\bibnamefont {Surrente}},
  \bibinfo {author} {\bibfnamefont {K.}~\bibnamefont {Galkowski}}, \bibinfo
  {author} {\bibfnamefont {A.}~\bibnamefont {Miyata}}, \bibinfo {author}
  {\bibfnamefont {O.}~\bibnamefont {Portugall}}, \bibinfo {author}
  {\bibfnamefont {R.~J.}\ \bibnamefont {Sutton}}, \bibinfo {author}
  {\bibfnamefont {A.~A.}\ \bibnamefont {Haghighirad}}, \bibinfo {author}
  {\bibfnamefont {H.~J.}\ \bibnamefont {Snaith}}, \bibinfo {author}
  {\bibfnamefont {D.~K.}\ \bibnamefont {Maude}}, \bibinfo {author}
  {\bibfnamefont {P.}~\bibnamefont {Plochocka}},\ and\ \bibinfo {author}
  {\bibfnamefont {R.~J.}\ \bibnamefont {Nicholas}},\ }\bibfield  {title}
  {\enquote {\bibinfo {title} {Impact of the halide cage on the electronic
  properties of fully inorganic cesium lead halide perovskites},}\ }\href
  {https://doi.org/10.1021/acsenergylett.7b00416} {\bibfield  {journal}
  {\bibinfo  {journal} {{ACS} Energy Lett.}\ }\textbf {\bibinfo {volume} {2}},\
  \bibinfo {pages} {1621--1627} (\bibinfo {year} {2017})}\BibitemShut {NoStop}%
\bibitem [{\citenamefont {Yu}\ \emph {et~al.}(2011)\citenamefont {Yu},
  \citenamefont {Chen}, \citenamefont {Wang}, \citenamefont {Pfenninger},
  \citenamefont {Vockic}, \citenamefont {Kenney},\ and\ \citenamefont
  {Shum}}]{Yu}%
  \BibitemOpen
  \bibfield  {author} {\bibinfo {author} {\bibfnamefont {C.}~\bibnamefont
  {Yu}}, \bibinfo {author} {\bibfnamefont {Z.}~\bibnamefont {Chen}}, \bibinfo
  {author} {\bibfnamefont {J.~J.}\ \bibnamefont {Wang}}, \bibinfo {author}
  {\bibfnamefont {W.}~\bibnamefont {Pfenninger}}, \bibinfo {author}
  {\bibfnamefont {N.}~\bibnamefont {Vockic}}, \bibinfo {author} {\bibfnamefont
  {J.~T.}\ \bibnamefont {Kenney}},\ and\ \bibinfo {author} {\bibfnamefont
  {K.}~\bibnamefont {Shum}},\ }\bibfield  {title} {\enquote {\bibinfo {title}
  {Temperature dependence of the band gap of perovskite semiconductor compound
  {CsSnI$_3$}},}\ }\href {https://doi.org/10.1063/1.3638699} {\bibfield
  {journal} {\bibinfo  {journal} {J. Appl. Phys.}\ }\textbf {\bibinfo {volume}
  {110}},\ \bibinfo {pages} {063526} (\bibinfo {year} {2011})}\BibitemShut
  {NoStop}%
\bibitem [{\citenamefont {Saran}\ \emph {et~al.}(2017)\citenamefont {Saran},
  \citenamefont {Heuer-Jungemann}, \citenamefont {Kanaras},\ and\ \citenamefont
  {Curry}}]{Saran}%
  \BibitemOpen
  \bibfield  {author} {\bibinfo {author} {\bibfnamefont {R.}~\bibnamefont
  {Saran}}, \bibinfo {author} {\bibfnamefont {A.}~\bibnamefont
  {Heuer-Jungemann}}, \bibinfo {author} {\bibfnamefont {A.~G.}\ \bibnamefont
  {Kanaras}},\ and\ \bibinfo {author} {\bibfnamefont {R.~J.}\ \bibnamefont
  {Curry}},\ }\bibfield  {title} {\enquote {\bibinfo {title} {Giant bandgap
  renormalization and exciton-phonon scattering in perovskite nanocrystals},}\
  }\href {https://doi.org/10.1002/adom.201700231} {\bibfield  {journal}
  {\bibinfo  {journal} {Adv. Opt. Mater.}\ }\textbf {\bibinfo {volume} {5}},\
  \bibinfo {pages} {1700231} (\bibinfo {year} {2017})}\BibitemShut {NoStop}%
\bibitem [{\citenamefont {Handa}\ \emph {et~al.}(2018)\citenamefont {Handa},
  \citenamefont {Aharen}, \citenamefont {Wakamiya},\ and\ \citenamefont
  {Kanemitsu}}]{Handa}%
  \BibitemOpen
  \bibfield  {author} {\bibinfo {author} {\bibfnamefont {T.}~\bibnamefont
  {Handa}}, \bibinfo {author} {\bibfnamefont {T.}~\bibnamefont {Aharen}},
  \bibinfo {author} {\bibfnamefont {A.}~\bibnamefont {Wakamiya}},\ and\
  \bibinfo {author} {\bibfnamefont {Y.}~\bibnamefont {Kanemitsu}},\ }\bibfield
  {title} {\enquote {\bibinfo {title} {Radiative recombination and
  electron-phonon coupling in lead-free {CH$_3$NH$_3$SnI$_3$} perovskite thin
  films},}\ }\href {https://doi.org/10.1103/physrevmaterials.2.075402}
  {\bibfield  {journal} {\bibinfo  {journal} {Phys. Rev. Mater.}\ }\textbf
  {\bibinfo {volume} {2}} (\bibinfo {year} {2018}),\
  10.1103/physrevmaterials.2.075402}\BibitemShut {NoStop}%
\bibitem [{\citenamefont {Gibson}(1952)}]{Gibson}%
  \BibitemOpen
  \bibfield  {author} {\bibinfo {author} {\bibfnamefont {A.~F.}\ \bibnamefont
  {Gibson}},\ }\bibfield  {title} {\enquote {\bibinfo {title} {The absorption
  spectra of single crystals of lead sulphide, selenide and telluride},}\
  }\href {https://doi.org/10.1088/0370-1301/65/5/309} {\bibfield  {journal}
  {\bibinfo  {journal} {Proc. Phys. Soc. London, Sect. B}\ }\textbf {\bibinfo
  {volume} {65}},\ \bibinfo {pages} {378--388} (\bibinfo {year}
  {1952})}\BibitemShut {NoStop}%
\bibitem [{\citenamefont {Gibbs}\ \emph {et~al.}(2013)\citenamefont {Gibbs},
  \citenamefont {Kim}, \citenamefont {Wang}, \citenamefont {White},
  \citenamefont {Drymiotis}, \citenamefont {Kaviany},\ and\ \citenamefont
  {Snyder}}]{Gibbs}%
  \BibitemOpen
  \bibfield  {author} {\bibinfo {author} {\bibfnamefont {Z.~M.}\ \bibnamefont
  {Gibbs}}, \bibinfo {author} {\bibfnamefont {H.}~\bibnamefont {Kim}}, \bibinfo
  {author} {\bibfnamefont {H.}~\bibnamefont {Wang}}, \bibinfo {author}
  {\bibfnamefont {R.~L.}\ \bibnamefont {White}}, \bibinfo {author}
  {\bibfnamefont {F.}~\bibnamefont {Drymiotis}}, \bibinfo {author}
  {\bibfnamefont {M.}~\bibnamefont {Kaviany}},\ and\ \bibinfo {author}
  {\bibfnamefont {G.~J.}\ \bibnamefont {Snyder}},\ }\bibfield  {title}
  {\enquote {\bibinfo {title} {Temperature dependent band gap in {P}b{X}
  ({X}{\hspace{0.167em}}={\hspace{0.167em}}{S}, {S}e, {T}e)},}\ }\href
  {https://doi.org/10.1063/1.4858195} {\bibfield  {journal} {\bibinfo
  {journal} {Appl. Phys. Lett.}\ }\textbf {\bibinfo {volume} {103}},\ \bibinfo
  {pages} {262109} (\bibinfo {year} {2013})}\BibitemShut {NoStop}%
\bibitem [{\citenamefont {Paul}\ and\ \citenamefont {Jones}(1953)}]{Paul}%
  \BibitemOpen
  \bibfield  {author} {\bibinfo {author} {\bibfnamefont {W.}~\bibnamefont
  {Paul}}\ and\ \bibinfo {author} {\bibfnamefont {R.~V.}\ \bibnamefont
  {Jones}},\ }\bibfield  {title} {\enquote {\bibinfo {title} {Absorption
  spectra of lead sulphide at different temperatures},}\ }\href
  {https://doi.org/10.1088/0370-1301/66/3/307} {\bibfield  {journal} {\bibinfo
  {journal} {Proc. Phys. Soc. London, Sect. B}\ }\textbf {\bibinfo {volume}
  {66}},\ \bibinfo {pages} {194--200} (\bibinfo {year} {1953})}\BibitemShut
  {NoStop}%
\bibitem [{\citenamefont {Tauber}, \citenamefont {Machonis},\ and\
  \citenamefont {Cadoff}(1966)}]{Tauber}%
  \BibitemOpen
  \bibfield  {author} {\bibinfo {author} {\bibfnamefont {R.~N.}\ \bibnamefont
  {Tauber}}, \bibinfo {author} {\bibfnamefont {A.~A.}\ \bibnamefont
  {Machonis}},\ and\ \bibinfo {author} {\bibfnamefont {I.~B.}\ \bibnamefont
  {Cadoff}},\ }\bibfield  {title} {\enquote {\bibinfo {title} {Thermal and
  optical energy gaps in {PbTe}},}\ }\href {https://doi.org/10.1063/1.1708150}
  {\bibfield  {journal} {\bibinfo  {journal} {J. Appl. Phys.}\ }\textbf
  {\bibinfo {volume} {37}},\ \bibinfo {pages} {4855--4860} (\bibinfo {year}
  {1966})}\BibitemShut {NoStop}%
\bibitem [{\citenamefont {Fabini}, \citenamefont {Seshadri},\ and\
  \citenamefont {Kanatzidis}(2020)}]{Fabini_MRS}%
  \BibitemOpen
  \bibfield  {author} {\bibinfo {author} {\bibfnamefont {D.~H.}\ \bibnamefont
  {Fabini}}, \bibinfo {author} {\bibfnamefont {R.}~\bibnamefont {Seshadri}},\
  and\ \bibinfo {author} {\bibfnamefont {M.~G.}\ \bibnamefont {Kanatzidis}},\
  }\bibfield  {title} {\enquote {\bibinfo {title} {The underappreciated lone
  pair in halide perovskites underpins their unusual properties},}\ }\href
  {https://doi.org/10.1557/mrs.2020.142} {\bibfield  {journal} {\bibinfo
  {journal} {{MRS} Bulletin}\ }\textbf {\bibinfo {volume} {45}},\ \bibinfo
  {pages} {467--477} (\bibinfo {year} {2020})}\BibitemShut {NoStop}%
\bibitem [{\citenamefont {Kontos}\ \emph {et~al.}(2018)\citenamefont {Kontos},
  \citenamefont {Kaltzoglou}, \citenamefont {Arfanis}, \citenamefont {McCall},
  \citenamefont {Stoumpos}, \citenamefont {Wessels}, \citenamefont {Falaras},\
  and\ \citenamefont {Kanatzidis}}]{Kontos}%
  \BibitemOpen
  \bibfield  {author} {\bibinfo {author} {\bibfnamefont {A.~G.}\ \bibnamefont
  {Kontos}}, \bibinfo {author} {\bibfnamefont {A.}~\bibnamefont {Kaltzoglou}},
  \bibinfo {author} {\bibfnamefont {M.~K.}\ \bibnamefont {Arfanis}}, \bibinfo
  {author} {\bibfnamefont {K.~M.}\ \bibnamefont {McCall}}, \bibinfo {author}
  {\bibfnamefont {C.~C.}\ \bibnamefont {Stoumpos}}, \bibinfo {author}
  {\bibfnamefont {B.~W.}\ \bibnamefont {Wessels}}, \bibinfo {author}
  {\bibfnamefont {P.}~\bibnamefont {Falaras}},\ and\ \bibinfo {author}
  {\bibfnamefont {M.~G.}\ \bibnamefont {Kanatzidis}},\ }\bibfield  {title}
  {\enquote {\bibinfo {title} {Dynamic disorder, band gap widening, and
  persistent near-{IR} photoluminescence up to at least 523 {K} in {ASnI$_3$}
  perovskites ({A = Cs$^{+}$, CH$_3$NH$_3^+$ and NH$_2$CH-NH$_2^+$})},}\ }\href
  {https://doi.org/10.1021/acs.jpcc.8b10218} {\bibfield  {journal} {\bibinfo
  {journal} {J. Phys. Chem. C}\ }\textbf {\bibinfo {volume} {122}},\ \bibinfo
  {pages} {26353--26361} (\bibinfo {year} {2018})}\BibitemShut {NoStop}%
\bibitem [{\citenamefont {Aeppli}\ \emph {et~al.}(2020)\citenamefont {Aeppli},
  \citenamefont {Balatsky}, \citenamefont {R{\o}nnow},\ and\ \citenamefont
  {Spaldin}}]{Aeppli}%
  \BibitemOpen
  \bibfield  {author} {\bibinfo {author} {\bibfnamefont {G.}~\bibnamefont
  {Aeppli}}, \bibinfo {author} {\bibfnamefont {A.~V.}\ \bibnamefont
  {Balatsky}}, \bibinfo {author} {\bibfnamefont {H.~M.}\ \bibnamefont
  {R{\o}nnow}},\ and\ \bibinfo {author} {\bibfnamefont {N.~A.}\ \bibnamefont
  {Spaldin}},\ }\bibfield  {title} {\enquote {\bibinfo {title} {Hidden,
  entangled and resonating order},}\ }\href
  {https://doi.org/10.1038/s41578-020-0207-z} {\bibfield  {journal} {\bibinfo
  {journal} {Nat. Rev. Mater.}\ }\textbf {\bibinfo {volume} {5}},\ \bibinfo
  {pages} {477--479} (\bibinfo {year} {2020})}\BibitemShut {NoStop}%
\bibitem [{\citenamefont {Lehn}(1970)}]{Lehn}%
  \BibitemOpen
  \bibfield  {author} {\bibinfo {author} {\bibfnamefont {J.~M.}\ \bibnamefont
  {Lehn}},\ }\bibfield  {title} {\enquote {\bibinfo {title} {Nitrogen
  inversion},}\ }in\ \href {https://doi.org/10.1007/bfb0050820} {\emph
  {\bibinfo {booktitle} {Fortschritte der Chemischen Forschung}}},\ Vol.\
  \bibinfo {volume} {15/3}\ (\bibinfo  {publisher} {Springer-Berlin},\ \bibinfo
  {year} {1970})\ pp.\ \bibinfo {pages} {311--377}\BibitemShut {NoStop}%
\bibitem [{\citenamefont {Silbey}\ and\ \citenamefont
  {Harris}(1984)}]{Silbey_Harris}%
  \BibitemOpen
  \bibfield  {author} {\bibinfo {author} {\bibfnamefont {R.}~\bibnamefont
  {Silbey}}\ and\ \bibinfo {author} {\bibfnamefont {R.~A.}\ \bibnamefont
  {Harris}},\ }\bibfield  {title} {\enquote {\bibinfo {title} {Variational
  calculation of the dynamics of a two level system interacting with a bath},}\
  }\href {https://doi.org/10.1063/1.447055} {\bibfield  {journal} {\bibinfo
  {journal} {J. Chem. Phys.}\ }\textbf {\bibinfo {volume} {80}},\ \bibinfo
  {pages} {2615--2617} (\bibinfo {year} {1984})}\BibitemShut {NoStop}%
\bibitem [{\citenamefont {Leggett}\ \emph {et~al.}(1995)\citenamefont
  {Leggett}, \citenamefont {Chakravarty}, \citenamefont {Dorsey}, \citenamefont
  {Fisher}, \citenamefont {Garg},\ and\ \citenamefont {Zwerger}}]{Leggett_RMP}%
  \BibitemOpen
  \bibfield  {author} {\bibinfo {author} {\bibfnamefont {A.~J.}\ \bibnamefont
  {Leggett}}, \bibinfo {author} {\bibfnamefont {S.}~\bibnamefont
  {Chakravarty}}, \bibinfo {author} {\bibfnamefont {A.~T.}\ \bibnamefont
  {Dorsey}}, \bibinfo {author} {\bibfnamefont {M.~P.~A.}\ \bibnamefont
  {Fisher}}, \bibinfo {author} {\bibfnamefont {A.}~\bibnamefont {Garg}},\ and\
  \bibinfo {author} {\bibfnamefont {W.}~\bibnamefont {Zwerger}},\ }\bibfield
  {title} {\enquote {\bibinfo {title} {Erratum: Dynamics of the dissipative
  two-state system},}\ }\href {https://doi.org/10.1103/revmodphys.67.725}
  {\bibfield  {journal} {\bibinfo  {journal} {Rev. Mod. Phys.}\ }\textbf
  {\bibinfo {volume} {67}},\ \bibinfo {pages} {725--726} (\bibinfo {year}
  {1995})}\BibitemShut {NoStop}%
\bibitem [{\citenamefont {Zurek}(1991)}]{Zurek}%
  \BibitemOpen
  \bibfield  {author} {\bibinfo {author} {\bibfnamefont {W.~H.}\ \bibnamefont
  {Zurek}},\ }\bibfield  {title} {\enquote {\bibinfo {title} {Decoherence and
  the transition from quantum to classical},}\ }\href
  {https://doi.org/10.1063/1.881293} {\bibfield  {journal} {\bibinfo  {journal}
  {Phys. Today}\ }\textbf {\bibinfo {volume} {44}},\ \bibinfo {pages} {36--44}
  (\bibinfo {year} {1991})}\BibitemShut {NoStop}%
\bibitem [{\citenamefont {Joos}(1996)}]{Joos1996}%
  \BibitemOpen
  \bibfield  {author} {\bibinfo {author} {\bibfnamefont {E.}~\bibnamefont
  {Joos}},\ }\href@noop {} {\emph {\bibinfo {title} {Decoherence and the
  Appearance of a Classical World in Quantum Theory}}},\ edited by\ \bibinfo
  {editor} {\bibfnamefont {D.}~\bibnamefont {Giulini}}, \bibinfo {editor}
  {\bibfnamefont {E.}~\bibnamefont {Joos}}, \bibinfo {editor} {\bibfnamefont
  {C.}~\bibnamefont {Kiefer}}, \bibinfo {editor} {\bibfnamefont
  {J.}~\bibnamefont {Kupsch}}, \bibinfo {editor} {\bibfnamefont {I.-O.}\
  \bibnamefont {Stamatescu}},\ and\ \bibinfo {editor} {\bibfnamefont {H.~D.}\
  \bibnamefont {Zeh}}\ (\bibinfo  {publisher} {Springer Berlin Heidelberg},\
  \bibinfo {year} {1996})\ pp.\ \bibinfo {pages} {35--136}\BibitemShut
  {NoStop}%
\bibitem [{\citenamefont {Simonius}(1978)}]{Simonius}%
  \BibitemOpen
  \bibfield  {author} {\bibinfo {author} {\bibfnamefont {M.}~\bibnamefont
  {Simonius}},\ }\bibfield  {title} {\enquote {\bibinfo {title} {Spontaneous
  symmetry breaking and blocking of metastable states},}\ }\href
  {https://doi.org/10.1103/physrevlett.40.980} {\bibfield  {journal} {\bibinfo
  {journal} {Phys. Rev. Lett.}\ }\textbf {\bibinfo {volume} {40}},\ \bibinfo
  {pages} {980--983} (\bibinfo {year} {1978})}\BibitemShut {NoStop}%
\bibitem [{\citenamefont {Wagner}(1986)}]{Wagner}%
  \BibitemOpen
  \bibfield  {author} {\bibinfo {author} {\bibfnamefont {M.}~\bibnamefont
  {Wagner}},\ }\bibfield  {title} {\enquote {\bibinfo {title} {{Unitary
  Transformations in Solid State Physics}},}\ }\href@noop {} {\bibfield
  {journal} {\bibinfo  {journal} {Mod. Probl. Condens. Matter Sci.}\ }\textbf
  {\bibinfo {volume} {15}},\ \bibinfo {pages} {1--343} (\bibinfo {year}
  {1986})}\BibitemShut {NoStop}%
\bibitem [{\citenamefont {Prager}\ and\ \citenamefont
  {Heidemann}(1997)}]{Prager-Heidemann}%
  \BibitemOpen
  \bibfield  {author} {\bibinfo {author} {\bibfnamefont {M.}~\bibnamefont
  {Prager}}\ and\ \bibinfo {author} {\bibfnamefont {A.}~\bibnamefont
  {Heidemann}},\ }\bibfield  {title} {\enquote {\bibinfo {title} {Rotational
  tunneling and neutron spectroscopy:~ a compilation},}\ }\href
  {https://doi.org/10.1021/cr9500848} {\bibfield  {journal} {\bibinfo
  {journal} {Chem. Rev.}\ }\textbf {\bibinfo {volume} {97}},\ \bibinfo {pages}
  {2933--2966} (\bibinfo {year} {1997})}\BibitemShut {NoStop}%
\bibitem [{\citenamefont {Ley}, \citenamefont {Gerbig},\ and\ \citenamefont
  {Schreiner}(2012)}]{Ley_Gerbig_Schreiner}%
  \BibitemOpen
  \bibfield  {author} {\bibinfo {author} {\bibfnamefont {D.}~\bibnamefont
  {Ley}}, \bibinfo {author} {\bibfnamefont {D.}~\bibnamefont {Gerbig}},\ and\
  \bibinfo {author} {\bibfnamefont {P.~R.}\ \bibnamefont {Schreiner}},\
  }\bibfield  {title} {\enquote {\bibinfo {title} {Tunnelling control of
  chemical reactions {\textendash} the organic chemist{\textquotesingle}s
  perspective},}\ }\href {https://doi.org/10.1039/c2ob07170c} {\bibfield
  {journal} {\bibinfo  {journal} {Org. Biomol. Chem.}\ }\textbf {\bibinfo
  {volume} {10}},\ \bibinfo {pages} {3781} (\bibinfo {year}
  {2012})}\BibitemShut {NoStop}%
\bibitem [{\citenamefont {Benderskii}, \citenamefont {Makarov},\ and\
  \citenamefont {Wight}(1994)}]{Benderskii}%
  \BibitemOpen
  \bibfield  {author} {\bibinfo {author} {\bibfnamefont {V.~A.}\ \bibnamefont
  {Benderskii}}, \bibinfo {author} {\bibfnamefont {D.~E.}\ \bibnamefont
  {Makarov}},\ and\ \bibinfo {author} {\bibfnamefont {C.~A.}\ \bibnamefont
  {Wight}},\ }\href
  {https://www.ebook.de/de/product/3661877/victor_a_benderskii_dmitrii_e_makarov_charles_a_wight_chemical_dynamics_at_low_temperatures.html}
  {\emph {\bibinfo {title} {Chemical Dynamics at Low Temperatures}}},\ Vol.\
  \bibinfo {volume} {LXXXVIII.}\ (\bibinfo  {publisher} {John Wiley},\ \bibinfo
  {year} {1994})\BibitemShut {NoStop}%
\bibitem [{\citenamefont {Sun}\ \emph {et~al.}(2020)\citenamefont {Sun},
  \citenamefont {Niu}, \citenamefont {Hermann}, \citenamefont {Moon},
  \citenamefont {Shulumba}, \citenamefont {Page}, \citenamefont {Zhao},
  \citenamefont {Thind}, \citenamefont {Mahalingam}, \citenamefont
  {Milam-Guerrero}, \citenamefont {Haiges}, \citenamefont {Mecklenburg},
  \citenamefont {Melot}, \citenamefont {Jho}, \citenamefont {Howe},
  \citenamefont {Mishra}, \citenamefont {Alatas}, \citenamefont {Winn},
  \citenamefont {Manley}, \citenamefont {Ravichandran},\ and\ \citenamefont
  {Minnich}}]{sun2020NatComm}%
  \BibitemOpen
  \bibfield  {author} {\bibinfo {author} {\bibfnamefont {B.}~\bibnamefont
  {Sun}}, \bibinfo {author} {\bibfnamefont {S.}~\bibnamefont {Niu}}, \bibinfo
  {author} {\bibfnamefont {R.~P.}\ \bibnamefont {Hermann}}, \bibinfo {author}
  {\bibfnamefont {J.}~\bibnamefont {Moon}}, \bibinfo {author} {\bibfnamefont
  {N.}~\bibnamefont {Shulumba}}, \bibinfo {author} {\bibfnamefont
  {K.}~\bibnamefont {Page}}, \bibinfo {author} {\bibfnamefont {B.}~\bibnamefont
  {Zhao}}, \bibinfo {author} {\bibfnamefont {A.~S.}\ \bibnamefont {Thind}},
  \bibinfo {author} {\bibfnamefont {K.}~\bibnamefont {Mahalingam}}, \bibinfo
  {author} {\bibfnamefont {J.}~\bibnamefont {Milam-Guerrero}}, \bibinfo
  {author} {\bibfnamefont {R.}~\bibnamefont {Haiges}}, \bibinfo {author}
  {\bibfnamefont {M.}~\bibnamefont {Mecklenburg}}, \bibinfo {author}
  {\bibfnamefont {B.~C.}\ \bibnamefont {Melot}}, \bibinfo {author}
  {\bibfnamefont {Y.-D.}\ \bibnamefont {Jho}}, \bibinfo {author} {\bibfnamefont
  {B.~M.}\ \bibnamefont {Howe}}, \bibinfo {author} {\bibfnamefont
  {R.}~\bibnamefont {Mishra}}, \bibinfo {author} {\bibfnamefont
  {A.}~\bibnamefont {Alatas}}, \bibinfo {author} {\bibfnamefont
  {B.}~\bibnamefont {Winn}}, \bibinfo {author} {\bibfnamefont {M.~E.}\
  \bibnamefont {Manley}}, \bibinfo {author} {\bibfnamefont {J.}~\bibnamefont
  {Ravichandran}},\ and\ \bibinfo {author} {\bibfnamefont {A.~J.}\ \bibnamefont
  {Minnich}},\ }\bibfield  {title} {\enquote {\bibinfo {title} {High frequency
  atomic tunneling yields ultralow and glass-like thermal conductivity in
  chalcogenide single crystals},}\ }\href
  {https://doi.org/https://doi.org/10.1038/s41467-020-19872-w} {\bibfield
  {journal} {\bibinfo  {journal} {Nature Communications}\ }\textbf {\bibinfo
  {volume} {11}},\ \bibinfo {pages} {1--9} (\bibinfo {year}
  {2020})}\BibitemShut {NoStop}%
\bibitem [{\citenamefont {Lautenschlager}\ \emph {et~al.}(1987)\citenamefont
  {Lautenschlager}, \citenamefont {Garriga}, \citenamefont {Logothetidis},\
  and\ \citenamefont {Cardona}}]{Lautenschlager_Cardona}%
  \BibitemOpen
  \bibfield  {author} {\bibinfo {author} {\bibfnamefont {P.}~\bibnamefont
  {Lautenschlager}}, \bibinfo {author} {\bibfnamefont {M.}~\bibnamefont
  {Garriga}}, \bibinfo {author} {\bibfnamefont {S.}~\bibnamefont
  {Logothetidis}},\ and\ \bibinfo {author} {\bibfnamefont {M.}~\bibnamefont
  {Cardona}},\ }\bibfield  {title} {\enquote {\bibinfo {title} {Interband
  critical points of {GaAs} and their temperature dependence},}\ }\href
  {https://doi.org/10.1103/physrevb.35.9174} {\bibfield  {journal} {\bibinfo
  {journal} {Phys. Rev. B}\ }\textbf {\bibinfo {volume} {35}},\ \bibinfo
  {pages} {9174--9189} (\bibinfo {year} {1987})}\BibitemShut {NoStop}%
\bibitem [{\citenamefont {Cheng}\ \emph {et~al.}(2020)\citenamefont {Cheng},
  \citenamefont {Qiao}, \citenamefont {Sheldon},\ and\ \citenamefont
  {Son}}]{Cheng}%
  \BibitemOpen
  \bibfield  {author} {\bibinfo {author} {\bibfnamefont {O.~H.-C.}\
  \bibnamefont {Cheng}}, \bibinfo {author} {\bibfnamefont {T.}~\bibnamefont
  {Qiao}}, \bibinfo {author} {\bibfnamefont {M.~T.}\ \bibnamefont {Sheldon}},\
  and\ \bibinfo {author} {\bibfnamefont {D.~H.}\ \bibnamefont {Son}},\
  }\bibfield  {title} {\enquote {\bibinfo {title} {Size-and
  temperature-dependent photoluminescence spectra of strongly confined
  {CsPbBr$_3$} quantum dots},}\ }\href {https://doi.org/10.1039/D0NR02711A}
  {\bibfield  {journal} {\bibinfo  {journal} {Nanoscale}\ }\textbf {\bibinfo
  {volume} {12}},\ \bibinfo {pages} {13113--13118} (\bibinfo {year}
  {2020})}\BibitemShut {NoStop}%
\bibitem [{\citenamefont {Soufiani}\ \emph {et~al.}(2015)\citenamefont
  {Soufiani}, \citenamefont {Huang}, \citenamefont {Reece}, \citenamefont
  {Sheng}, \citenamefont {Ho-Baillie},\ and\ \citenamefont {Green}}]{Soufiani}%
  \BibitemOpen
  \bibfield  {author} {\bibinfo {author} {\bibfnamefont {A.~M.}\ \bibnamefont
  {Soufiani}}, \bibinfo {author} {\bibfnamefont {F.}~\bibnamefont {Huang}},
  \bibinfo {author} {\bibfnamefont {P.}~\bibnamefont {Reece}}, \bibinfo
  {author} {\bibfnamefont {R.}~\bibnamefont {Sheng}}, \bibinfo {author}
  {\bibfnamefont {A.}~\bibnamefont {Ho-Baillie}},\ and\ \bibinfo {author}
  {\bibfnamefont {M.~A.}\ \bibnamefont {Green}},\ }\bibfield  {title} {\enquote
  {\bibinfo {title} {Polaronic exciton binding energy in iodide and bromide
  organic-inorganic lead halide perovskites},}\ }\href
  {https://doi.org/10.1063/1.4936418} {\bibfield  {journal} {\bibinfo
  {journal} {Applied Physics Letters}\ }\textbf {\bibinfo {volume} {107}},\
  \bibinfo {pages} {231902} (\bibinfo {year} {2015})},\ \Eprint
  {https://arxiv.org/abs/https://doi.org/10.1063/1.4936418}
  {https://doi.org/10.1063/1.4936418} \BibitemShut {NoStop}%
\bibitem [{\citenamefont {Tilchin}\ \emph {et~al.}(2016)\citenamefont
  {Tilchin}, \citenamefont {Dirin}, \citenamefont {Maikov}, \citenamefont
  {Sashchiuk}, \citenamefont {Kovalenko},\ and\ \citenamefont
  {Lifshitz}}]{Tilchin}%
  \BibitemOpen
  \bibfield  {author} {\bibinfo {author} {\bibfnamefont {J.}~\bibnamefont
  {Tilchin}}, \bibinfo {author} {\bibfnamefont {D.~N.}\ \bibnamefont {Dirin}},
  \bibinfo {author} {\bibfnamefont {G.~I.}\ \bibnamefont {Maikov}}, \bibinfo
  {author} {\bibfnamefont {A.}~\bibnamefont {Sashchiuk}}, \bibinfo {author}
  {\bibfnamefont {M.~V.}\ \bibnamefont {Kovalenko}},\ and\ \bibinfo {author}
  {\bibfnamefont {E.}~\bibnamefont {Lifshitz}},\ }\bibfield  {title} {\enquote
  {\bibinfo {title} {Hydrogen-like wannier-mott excitons in single crystal of
  methylammonium lead bromide perovskite},}\ }\href
  {https://doi.org/10.1021/acsnano.6b02734} {\bibfield  {journal} {\bibinfo
  {journal} {ACS Nano}\ }\textbf {\bibinfo {volume} {10}},\ \bibinfo {pages}
  {6363--6371} (\bibinfo {year} {2016})},\ \bibinfo {note} {pMID: 27249335},\
  \Eprint {https://arxiv.org/abs/https://doi.org/10.1021/acsnano.6b02734}
  {https://doi.org/10.1021/acsnano.6b02734} \BibitemShut {NoStop}%
\bibitem [{\citenamefont {Ruf}\ \emph {et~al.}(2019)\citenamefont {Ruf},
  \citenamefont {Ayg\''uler}, \citenamefont {Giesbrecht}, \citenamefont
  {Rendenbach}, \citenamefont {Magin}, \citenamefont {Docampo}, \citenamefont
  {Kalt},\ and\ \citenamefont {Hetterich}}]{Ruf}%
  \BibitemOpen
  \bibfield  {author} {\bibinfo {author} {\bibfnamefont {F.}~\bibnamefont
  {Ruf}}, \bibinfo {author} {\bibfnamefont {M.~F.}\ \bibnamefont {Ayg\''uler}},
  \bibinfo {author} {\bibfnamefont {N.}~\bibnamefont {Giesbrecht}}, \bibinfo
  {author} {\bibfnamefont {B.}~\bibnamefont {Rendenbach}}, \bibinfo {author}
  {\bibfnamefont {A.}~\bibnamefont {Magin}}, \bibinfo {author} {\bibfnamefont
  {P.}~\bibnamefont {Docampo}}, \bibinfo {author} {\bibfnamefont
  {H.}~\bibnamefont {Kalt}},\ and\ \bibinfo {author} {\bibfnamefont
  {M.}~\bibnamefont {Hetterich}},\ }\bibfield  {title} {\enquote {\bibinfo
  {title} {Temperature-dependent studies of exciton binding energy and
  phase-transition suppression in {(Cs,FA,MA)Pb(I,Br)$_3$} perovskites},}\
  }\href {https://doi.org/10.1063/1.5083792} {\bibfield  {journal} {\bibinfo
  {journal} {APL Materials}\ }\textbf {\bibinfo {volume} {7}},\ \bibinfo
  {pages} {031113} (\bibinfo {year} {2019})},\ \Eprint
  {https://arxiv.org/abs/https://doi.org/10.1063/1.5083792}
  {https://doi.org/10.1063/1.5083792} \BibitemShut {NoStop}%
\bibitem [{\citenamefont {Wang}\ \emph {et~al.}(2019)\citenamefont {Wang},
  \citenamefont {Ma}, \citenamefont {Li}, \citenamefont {Wang}, \citenamefont
  {Wang}, \citenamefont {Shen}, \citenamefont {Li}, \citenamefont {Wang},
  \citenamefont {Luo},\ and\ \citenamefont {Li}}]{Wang}%
  \BibitemOpen
  \bibfield  {author} {\bibinfo {author} {\bibfnamefont {S.}~\bibnamefont
  {Wang}}, \bibinfo {author} {\bibfnamefont {J.}~\bibnamefont {Ma}}, \bibinfo
  {author} {\bibfnamefont {W.}~\bibnamefont {Li}}, \bibinfo {author}
  {\bibfnamefont {J.}~\bibnamefont {Wang}}, \bibinfo {author} {\bibfnamefont
  {H.}~\bibnamefont {Wang}}, \bibinfo {author} {\bibfnamefont {H.}~\bibnamefont
  {Shen}}, \bibinfo {author} {\bibfnamefont {J.}~\bibnamefont {Li}}, \bibinfo
  {author} {\bibfnamefont {J.}~\bibnamefont {Wang}}, \bibinfo {author}
  {\bibfnamefont {H.}~\bibnamefont {Luo}},\ and\ \bibinfo {author}
  {\bibfnamefont {D.}~\bibnamefont {Li}},\ }\bibfield  {title} {\enquote
  {\bibinfo {title} {Temperature-dependent band gap in two-dimensional
  perovskites: Thermal expansion interaction and electron-phonon
  interaction},}\ }\href {https://doi.org/10.1021/acs.jpclett.9b01011}
  {\bibfield  {journal} {\bibinfo  {journal} {The Journal of Physical Chemistry
  Letters}\ }\textbf {\bibinfo {volume} {10}},\ \bibinfo {pages} {2546--2553}
  (\bibinfo {year} {2019})},\ \Eprint
  {https://arxiv.org/abs/https://doi.org/10.1021/acs.jpclett.9b01011}
  {https://doi.org/10.1021/acs.jpclett.9b01011} \BibitemShut {NoStop}%
\end{thebibliography}%
\begin{widetext}

\vspace{1cm}
\noindent
{\Large
{\bf Supplementary Material}
}

\section{Calculation of the renormalized tunnel splitting\cite{Wagner} $\Delta_r(T)$}

The Hamiltonian, Eq.  (\ref{eqn: Hamiltonian}) [main text], goes by various names in different contexts (e.g. the Jaynes-Cummings model for cavity electrodynamics or the Shore-Sander model\cite{Wagner} for polaronic self-trapping of the exciton) and generalizes to the archetypal Caldeira-Leggett model for quantum dissipation\cite{Leggett_RMP} when the bosonic bath is assumed to span a continuum of frequencies. Since our problem is essentially that of vibronic coupling, a discrete Einstein phonon with a characteristic frequency $\Omega_0$ seems sufficient to describe the essential physics.

Here we give the detailed derivation for the temperature dependent renormalization of the tunnel splitting $\Delta_r(T)$ [Eq. 2 (main text)],
\begin{equation}\label{Eqn:SplittingFinal}
 \Delta_r(T) \approx \Delta_0\exp[-2\left(\frac{\kappa}{\hbar\Omega_0}\right)^2\coth(\frac{\hbar\Omega_0}{2k_BT})]\nonumber
\end{equation}
starting from the spin-boson Hamiltonian [Eq. 1 (main text)]
\begin{eqnarray}
H = \Delta_0 \hat{\sigma}_x + \hbar \Omega_0 \left(\hat{a}^\dagger \hat{a}+\frac{1}{2}\right) +\kappa \hat{\sigma}_z (\hat{a}+\hat{a}^\dagger), \label{main_hamiltonian}
\end{eqnarray}
with a bare tunnel splitting $\Delta_0$. The is a standard calculation, although it is slightly involved.

\subsection{Variational Analysis}
We begin by writing the above Hamiltonian as $H = H_0 + H_B + H_I$, where $H_0$, $H_B$ and $H_I$ represent the three terms in Eq. (\ref{main_hamiltonian}) respectively. We diagonalize this Hamiltonian using the polaronic unitary transformation with the unitary matrix $U$,
\begin{eqnarray}
U = \exp[-\hat{\sigma}_z\frac{f}{\hbar\Omega_0}(\hat{a}-\hat{a}^\dagger)] \label{unitary_transform}
\end{eqnarray}
where $f$ is the variational parameter. The transformed Hamiltonian will then become as follows,
\begin{eqnarray}
&\widetilde{H} = UHU^\dagger = \widetilde{H}_0 + \widetilde{H}_B + \widetilde{H}_I \label{transformed_hamiltonian}
\end{eqnarray}

We first evaluate the $\widetilde{H}_0$ term in Eq. (\ref{transformed_hamiltonian}). Let us rewrite the Eq. (\ref{unitary_transform}) as $U = \exp(-\hat{\sigma}_z\hat{\lambda})$, where $\hat{\lambda} = \frac{f}{\hbar\Omega_0}(\hat{a}-\hat{a}^\dagger)$, then one can write $\widetilde{H}_0$ as following,
\begin{eqnarray}
\widetilde{H}_0 = \Delta_0 \cosh(2\hat{\lambda})\hat{\sigma}_x-\imath\Delta_0 \sinh(2\hat{\lambda})\hat{\sigma}_y \label{first_term}
\end{eqnarray}
where we have used the fact that $\exp(-\hat{\sigma}_z\hat{\lambda})=\cosh\hat{\lambda}I-\sinh\hat{\lambda}\hat{\sigma_z}$ and $[\hat{\sigma}_z,\hat{\sigma}_x]=2\imath\hat{\sigma}_y$.

Next, we evaluate the $\widetilde{H}_I$ term in Eq. (\ref{transformed_hamiltonian}). We rewrite the Eq. (\ref{unitary_transform}) as $U=\exp(-S)$, where $S = \hat{\sigma}_z\hat{\lambda}$. Using the Baker-Campbell-Hausdorff (BCH) expansion, we can write $\widetilde{H}_I$ as following,
\begin{eqnarray}
\widetilde{H_I} &=&\exp(-S)H_I\exp(S) \notag \\
&=& H_I + [S,H_I] + \frac{1}{2!}[S,[S,H_I]] + ... \notag \\
&=& \kappa\hat{\sigma}_z(\hat{a}+\hat{a}^\dagger)-2\frac{\kappa f}{\hbar\Omega_0} \label{second_term}
\end{eqnarray}
where $[S,H_I] = 2\kappa f/\hbar\Omega_0 $ and higher order terms do not contribute.

Similarly, we evaluate the $\widetilde{H}_B$ term in Eq. (\ref{transformed_hamiltonian}). Using the BCH expansion, we can write $\widetilde{H}_B$ as following,
\begin{eqnarray}
\widetilde{H}_B &=& \exp(-S)H_B\exp(S) \notag \\
&=& H_B + [S,H_B] + \frac{1}{2!}[S,[S,H_B]] + ... \notag\\
&=& \hbar\Omega_0\left(\hat{a}^\dagger\hat{a}+\frac{1}{2}\right)-\hat{\sigma}_zf(\hat{a}+\hat{a}^\dagger)+\frac{f^2}{\hbar\Omega_0} \label{third_term}
\end{eqnarray}
where $[S,H_B] = -\hat{\sigma}_z f (\hat{a}+\hat{a}^\dagger)$ and $[S,[S,H_B]] = 2f^2/\hbar\Omega_0$, while all higher order terms do not contribute.

Next, we do the thermal averaging of the transformed Hamiltonian $\widetilde{H}$ by tracing out the bath degrees of freedom and obtain the effective spin Hamiltonian $H_S$ (the details of the derivation is given in the next subsection), which can be written as following,
\begin{eqnarray}
H_S &=& \Delta_r \hat{\sigma}_x + \frac{f^2}{\hbar\Omega_0}-2\frac{\kappa f}{\hbar\Omega_0} \label{spin_hamiltonian} \\
\text{where}\,\,\,\, \Delta_r(f,T) &=& \Delta_0\exp[-2\left(\frac{f}{\hbar\Omega_0}\right)^2\coth(\frac{\hbar\Omega_0}{2k_BT})] \notag
\end{eqnarray}

Since the effective spin Hamiltonian describes a two level system with bonding $\ket{\Phi_-}$ and anti-bonding $\ket{\Phi_+}$ wave functions, we can compute the energies $\epsilon_{\mp} = \langle \Phi_{\mp}|H_S|\Phi_{\mp}\rangle$, which can be written as following,
\begin{eqnarray}
\epsilon_{\mp} = \frac{f^2}{\hbar\Omega_0}-2\frac{\kappa f}{\hbar\Omega_0}\mp\Delta_r \label{TLS_energies}
\end{eqnarray}
Next, we construct the partition function $Z_S$ for this two level system,
\begin{eqnarray}
Z_S &=& \sum_{\epsilon} e^{-\epsilon/k_BT} \notag \\
&=& e^{-\epsilon_+/k_BT}+e^{-\epsilon_-/k_BT} \notag \\
&=& 2\cosh{\left(\frac{\Delta_r}{k_BT}\right)}\exp[-\frac{1}{k_BT}\left(\frac{f^2}{\hbar\Omega_0}-2\frac{\kappa f}{\hbar\Omega_0}\right)] \label{spin_partition_function}
\end{eqnarray}
and write the free energy $F$ as following,
\begin{eqnarray}
F &=& -k_BT\ln(Z_S) \notag\\
&=& -k_BT\ln(2)-k_BT\ln\left(\cosh{\left(\frac{\Delta_r}{k_BT}\right)}\right)+ \frac{f^2}{\hbar\Omega_0}-2\frac{\kappa f}{\hbar\Omega_0} \label{free_energy}
\end{eqnarray}
Minimizing the free energy w.r.t $f$ ($\partial F/\partial f =0 )$ gives the value of $f$ as following,
\begin{eqnarray}
f = \frac{\kappa}{1+\frac{2\Delta_r}{\hbar\Omega_0}\coth(\frac{\hbar\Omega_0}{2k_BT})\tanh(\frac{\Delta_r}{k_BT})}
\end{eqnarray}
Since  $\Delta_0 <<k_BT$ for our systems, $f=\kappa$ is a good approximation and equation (2) [main text] follows.
\subsection{Thermal expectation}
The transformed Hamiltonian can be written as following,
\begin{eqnarray}
\widetilde{H} = \frac{f^2}{\hbar\Omega_0}-2\frac{\kappa f}{\hbar\Omega_0} + \hbar\Omega_0\left(\hat{a}^\dagger\hat{a}+\frac{1}{2}\right) + \hat{\sigma}_z(\kappa-f)(\hat{a}+\hat{a}^\dagger)+\Delta_0(\cosh{2\hat{\lambda}}\hat{\sigma}_x-\imath\sinh{2\hat{\lambda}}\hat{\sigma}_y) \label{full_transformed_hamiltonian}
\end{eqnarray}
The thermal expectation of any operator $\hat{O}$ is given by $\langle \hat{O}  \rangle_{th} = \frac{1}{Z_B}\int d\alpha d\alpha^\star \bra{\alpha} \exp\left[-{\hbar \Omega_0\over k_BT} (a^\dagger a)\right]\hat{O}\ket{\alpha}$, where $\ket{\alpha}$ represents the bath coherent state ($\hat{a}\ket{\alpha}=\alpha\ket{\alpha}$) and $Z_B = \int d\alpha d\alpha^\star \langle \alpha | \exp(-\frac{\hbar\Omega_0}{k_BT}\hat{a}^\dagger\hat{a})|\alpha \rangle$. We first evaluate $Z_B$ as following,
\begin{eqnarray}
Z_B &=& \int d\alpha d\alpha^\star \exp[-\frac{\hbar\Omega_0}{k_BT}(\alpha^\star \alpha)] \notag \\
&=& \int d\Re{\alpha}d\Im{\alpha}\exp[-\frac{\hbar\Omega_0}{k_BT}\left(\Re^2\{\alpha\}+\Im^2\{\alpha\}\right)], \,\,\,\, \text{where} \,\,\,\, \alpha = \Re{\alpha}+\imath\Im{\alpha}\notag \\
&=& \left(\int d\Re{\alpha}\exp[-\frac{\hbar\Omega_0}{k_BT}\Re^2{\alpha}]\right)\left(\int d\Im{\alpha}\exp[-\frac{\hbar\Omega_0}{k_BT}\Im^2{\alpha}]\right) \notag \\
&=& \pi \frac{k_BT}{\hbar \Omega_0}
\end{eqnarray}

The only important terms in Eq. (\ref{full_transformed_hamiltonian}) for which we need to evaluate the thermal expectation is the last term as the first term is a constant, second term would give a constant independent of $f$ and third term contributes to zero. Since the last term can be written as a $2\times2$ matrix in the following way,
\begin{eqnarray}
\Delta_0\begin{bmatrix}
0 & \exp(-2\hat{\lambda}) \\
\exp(2\hat{\lambda}) & 0
\end{bmatrix} \label{matrix_form}
\end{eqnarray}
we only need to evaluate $\langle \exp(-2\hat{\lambda}) \rangle_{th}$ as follows,
\begin{eqnarray}
\langle \exp(-2\hat{\lambda}) \rangle_{th} = \frac{1}{Z_B} \int d\alpha d\alpha^\star \bra{\alpha}\exp[-\frac{\hbar\Omega_0}{k_BT}(\hat{a}^\dagger\hat{a})]\exp[-\frac{2f}{\hbar\Omega_0}(\hat{a}-\hat{a}^\dagger)]\ket{\alpha} \label{last_term_thermal_avg}
\end{eqnarray}
We can next use the following formula,
\begin{eqnarray}
\exp(X)\exp(Y)=\exp(Z) = \exp(X + Y + \frac{1}{2!}[X,Y] + \frac{1}{12}([X,[X,Y]-[Y,[X,Y]]) +...) \notag
\end{eqnarray}
where $X = -\frac{\hbar\Omega_0}{k_BT}\hat{a}^\dagger \hat{a}$ and $Y = -\frac{2f}{\hbar\Omega_0}(\hat{a}-\hat{a}^\dagger)$. One can then evaluate $\bra{\alpha}Z\ket{\alpha}$ as following,
\begin{eqnarray}
\bra{\alpha}Z\ket{\alpha} &=& -\frac{\hbar\Omega_0}{k_BT}\alpha^\star \alpha-\frac{2f}{\hbar\Omega_0}(\alpha-\alpha^\star)+\frac{f}{k_BT}(\alpha+\alpha^\star)-\frac{2f^2}{3k_BT\hbar\Omega_0}-\frac{f\hbar\Omega_0}{6(k_BT)^2}+... \notag \\
&=& -4\frac{\imath f}{\hbar\Omega_0}\left[1+\frac{1}{12}\left(\frac{\hbar\Omega_0}{k_BT}\right)^2\right] \Im{\alpha} - \frac{\hbar\Omega_0}{k_BT}\left(\Re^2\{\alpha\}+\Im^2\{\alpha\}\right)+\frac{2f}{k_BT}\Re{\alpha}-\frac{2f^2}{3k_BT\hbar\Omega_0}+... \notag\\
&=& -\frac{\hbar\Omega_0}{k_BT}\left[\Im{\alpha}+\frac{2\imath f}{\hbar\Omega_0}\left(1+\frac{1}{12}\left(\frac{\hbar\Omega_0}{k_BT}\right)^2\right)\right]^2-\frac{\hbar\Omega_0}{k_BT}\left[\Re{\alpha}-\frac{f}{\hbar\Omega_0}\right]^2\notag\\
&&-2\left(\frac{f}{\hbar\Omega_0}\right)^2\left[\frac{2k_BT}{\hbar\Omega_0}+\frac{\hbar\Omega_0}{6k_BT}+O\left(\left(\frac{\hbar\Omega_0}{k_BT}\right)^3\right)\right]\notag\\
&=&-\frac{\hbar\Omega_0}{k_BT}\left[\Im{\alpha}+\frac{2\imath f}{\hbar\Omega_0}\left(1+\frac{1}{12}\left(\frac{\hbar\Omega_0}{k_BT}\right)^2\right)\right]^2-\frac{\hbar\Omega_0}{k_BT}\left[\Re{\alpha}-\frac{f}{\hbar\Omega_0}\right]^2-2\left(\frac{f}{\hbar\Omega_0}\right)^2\coth{\left(\frac{\hbar\Omega_0}{2k_BT}\right)}
\end{eqnarray}
where we have used the expansion, $\coth{\left(\frac{\hbar\Omega_0}{2k_BT}\right)}=\left[\frac{2k_BT}{\hbar\Omega_0}+\frac{\hbar\Omega_0}{6k_BT}+O\left(\left(\frac{\hbar\Omega_0}{k_BT}\right)^3\right)\right]$. Continuing from Eq. (\ref{last_term_thermal_avg}), we have the following,
\begin{eqnarray}
\langle \exp(-2\hat{\lambda}) \rangle_{th} &=& \frac{e^{-2\left(\frac{f}{\hbar\Omega_0}\right)^2\coth{\left(\frac{\hbar\Omega_0}{2k_BT}\right)}}}{Z_B}\left(\int d\Im{\alpha}e^{-\frac{\hbar\Omega_0}{k_BT}\left[\Im{\alpha}+\frac{2\imath f}{\hbar\Omega_0}\left(1+\frac{1}{12}\left(\frac{\hbar\Omega_0}{k_BT}\right)^2\right)\right]^2} \right)\left(\int d\Re{\alpha} e^{-\frac{\hbar\Omega_0}{k_BT}\left[\Re{\alpha}-\frac{f}{\hbar\Omega_0}\right]^2} \right)\notag\\
&=&\exp[-2\left(\frac{f}{\hbar\Omega_0}\right)^2\coth{\left(\frac{\hbar\Omega_0}{2k_BT}\right)}] \label{thermal_factor}
\end{eqnarray}

Similarly, $\langle \exp(2\hat{\lambda}) \rangle_{th}$ will have the same value as the sign change would not affect the expression in Eq. (\ref{thermal_factor}). Therefore the matrix from Eq. (\ref{matrix_form}) can be written as follows,
\begin{eqnarray}
\Delta_0\begin{bmatrix}
0 & \exp[-2\left(\frac{f}{\hbar\Omega_0}\right)^2\coth{\left(\frac{\hbar\Omega_0}{2k_BT}\right)}] \\
\exp[-2\left(\frac{f}{\hbar\Omega_0}\right)^2\coth{\left(\frac{\hbar\Omega_0}{2k_BT}\right)}] & 0
\end{bmatrix} &=& \Delta_0 \exp[-2\left(\frac{f}{\hbar\Omega_0}\right)^2\coth{\left(\frac{\hbar\Omega_0}{2k_BT}\right)}]\hat{\sigma}_x \notag\\
&=&\Delta_r\hat{\sigma}_x
\end{eqnarray}

Thus, the thermal expectation of Eq. (\ref{full_transformed_hamiltonian}) will give Eq. (\ref{spin_hamiltonian}).

\section{Additional data on the temperature dependence of the bandgap in halide perovskites}
Figure 1 (supplement) shows some additional fits to the temperature dependence of the bandgap with the proposed formula

\begin{equation} \tag*{[Eq. 4 (main text)]}
E_g(T)=E_g^\infty-\Xi_0\exp[-\lbrace\exp(T_0/T)-1\rbrace ^{-1}],
\end{equation}
for a few other lead-containing halide perovskites besides those shown in the main text. The temperature dependence of excitonic bandgap of both inorganic and hybrid variants seems to be captured by the above equation. Information regarding the references and the extracted parameter values are recorded in Table 1 (supplement).
\setcounter{figure}{0}
\begin{figure}[h]
	\includegraphics[scale=0.37]{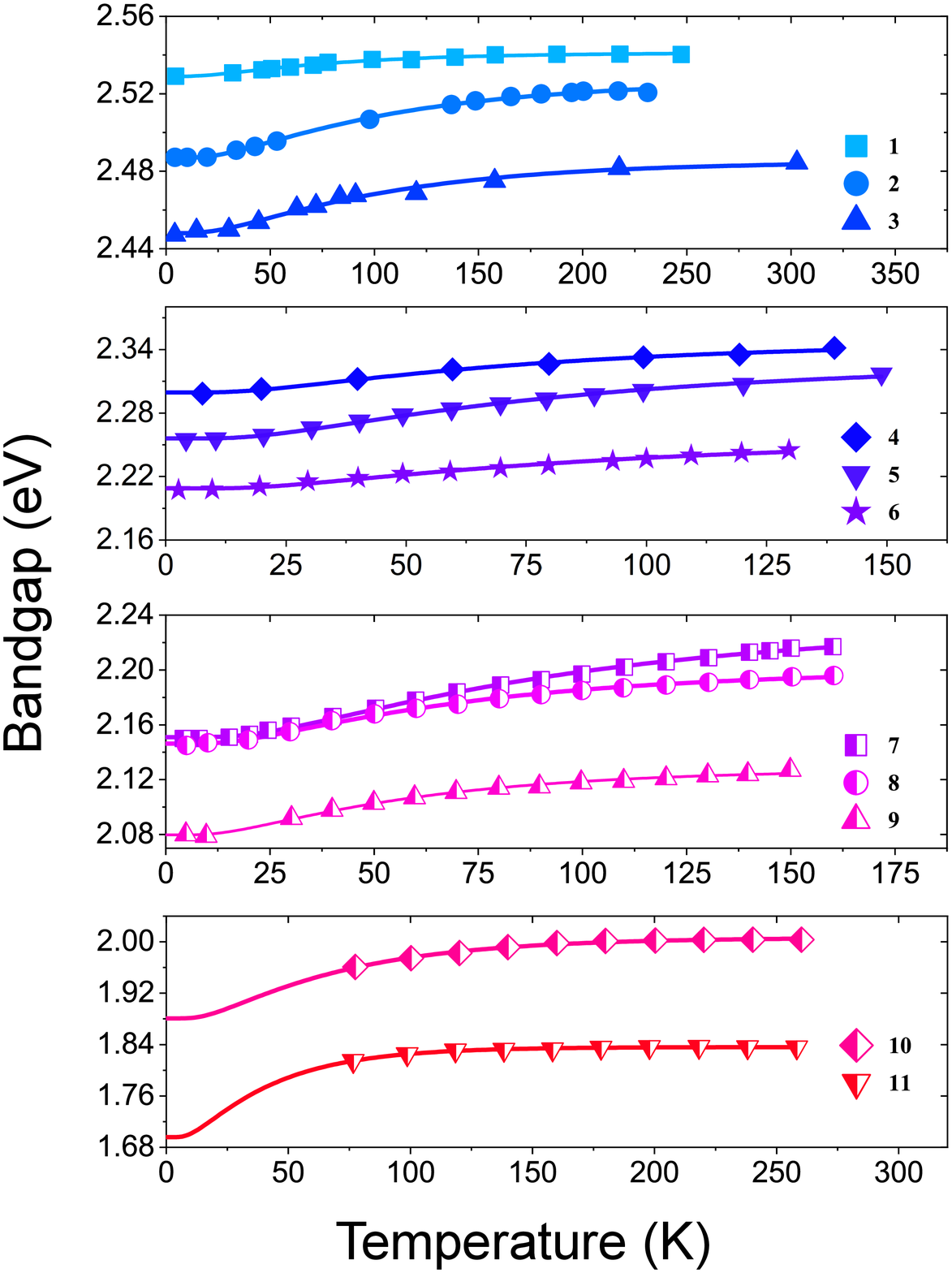}
	\centering
	\caption{(supplement): The scatter plots show the variation in bandgap of a few other lead-containing halide perovskite materials. The solid lines depict the fitting of those data with Eq. ~\ref{Eqn:Gap} above [corresponding to Eq. 4 (main text)]. Sample details and the extracted fitting parameters are listed in Table 1.}
\end{figure}
\begin{table*}[h]
	\centering
	\begin{tabular}{c c c c c c}
		\hline		
		\textbf{Sr. No.} & \textbf{Ref.} & \textbf{Sample} & \textbf{$E_g^\infty$(eV)} & \textbf{$\Xi_0$(meV)} & \textbf{T$_0$(K)}\\
		\hline
		 &  &  & &  &   \\
		1 & Cheng\cite{Cheng} & CsPbBr$_3$, 4.7nm QD & 2.541 & 12.16 & 62.87  \\
		2 & Cheng\cite{Cheng} & CsPbBr$_3$, 5.3nm QD & 2.526 & 39.50 & 83.93  \\
		3 & Cheng\cite{Cheng} & CsPbBr$_3$, 6.3nm QD & 2.486 & 36.94 & 82.11  \\
		4 & Soufiani\cite{Soufiani} & MAPbBr$_3$ & 2.347 & 47.11 & 58.71  \\
		5 & Tilchin\cite{Tilchin} & MAPbBr$_3$ & 2.326 & 69.56 & 64.63  \\
		6 & Yang\cite{Yang} & CsPbI$_2$Br & 2.254 & 45.43 & 68.58  \\
		7 & Ruf\cite{Ruf} & MAPbI$_3$ & 2.232 & 80.95 & 75.16  \\
		8 & Ruf\cite{Ruf} & Cs$_{0.1}$FA$_{0.765}$MA$_{0.135}$Pb(I$_{0.765}$Br$_{0.235}$)$_3$ & 2.200 & 53.53 & 56.24  \\
		9 & Ruf\cite{Ruf} & Cs$_{0.05}$(FA$_{0.83}$MA$_{0.17}$)$_{0.95}$Pb(I$_{0.83}$Br$_{0.17}$)$_3$ & 2.218 & 47.79 & 46.60  \\
		10 & Wang\cite{Wang} & (iso-BA)$_2$MA$_{n-1}$Pb$_n$I$_{3n+1}$, n=3 & 2.006 & 125.73 & 54.00  \\
		11 & Wang\cite{Wang} & (n-BA)$_2$MA$_{n-1}$Pb$_n$I$_{3n+1}$, n=5 & 1.836 & 140.29 & 32.71  \\
		&  &  & &  &   \\
		\hline
	\end{tabular}
	\caption{(supplement): Information regarding the plots in Fig.~1 (supplement) and extracted fitting parameters.}
\end{table*}

\end{widetext}

\end{document}